\documentclass{PoS}

\usepackage{lmodern}
\usepackage[T1]{fontenc}
\usepackage{amsmath,amssymb,amsfonts}

\usepackage{graphicx}
\usepackage{overpic}

\title{Janus and J-fold solutions in type IIB supergravity}

\ShortTitle{IIB Janus and J-folds}

\author{\speaker{Fri{\dh}rik Freyr Gautason}\\
        Instituut voor Theoretische Fysica, KU Leuven\\
		Celestijnenlaan 200D, 3001 Leuven, Belgium\\
		University of Iceland, Science Institute\\
		Dunhaga 3, 107 Reykjav{\'i}k, Iceland\\
        E-mail: \email{ffg@kuleuven.be}}

\abstract{I discuss recent supergravity constructions of type IIB holographic interfaces in four-dimensional ${\cal N}=1$ and ${\cal N}=4$ field theories. I explain how each holographic interface can be compactified on a circle with an $\SL(2,{\Z})$ monodromy leading to a novel AdS$_4$ supergravity solution. These AdS$_4$ backgrounds are argued to be dual to so-called J-fold superconformal field theories in three dimensions.}

\FullConference{Corfu Summer Institute 2019 "School and Workshops on Elementary Particle Physics and Gravity" (CORFU2019)\\
		31 August - 25 September 2019\\
		Corfu, Greece}

\usepackage{enumerate}
\numberwithin{equation}{section}

\newcommand{\dd}{\mathrm{d}}

\newcommand{\e}{\mathrm{e}}

\newcommand{\be}{\begin{equation}}
\newcommand{\ee}{\end{equation}}
\newcommand{\bea}{\begin{eqnarray}}
\newcommand{\eea}{\end{eqnarray}}

\newcommand{\f}[2]{\frac{#1}{#2}}
\newcommand{\R}{\mathbf{R}}
\newcommand{\Z}{\mathbf{Z}}
\newcommand{\N}{{\cal N}}

\renewcommand{\Re}{\text{Re}}
\renewcommand{\Im}{\text{Im}}

\newcommand{\OSp}{\mathrm{OSp}}

\newcommand{\U}{\text{U}}
\newcommand{\SU}{\text{SU}}
\newcommand{\SO}{\text{SO}}

\newcommand{\USp}{\text{USp}}
\newcommand{\SL}{\text{SL}}

\begin{document}

\section{Introduction}
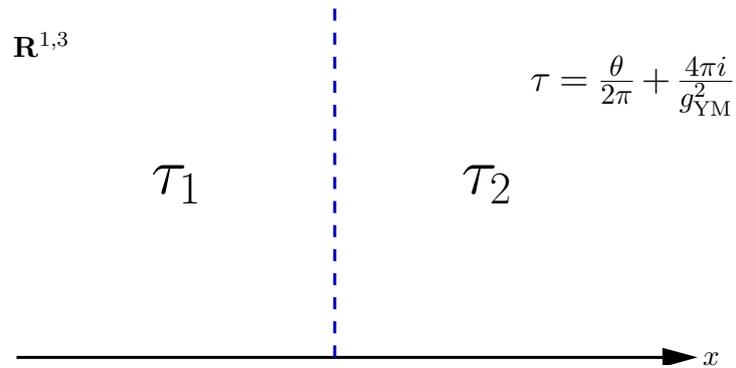
\begin{figure}\centering
\begin{overpic}[width=0.6\textwidth]{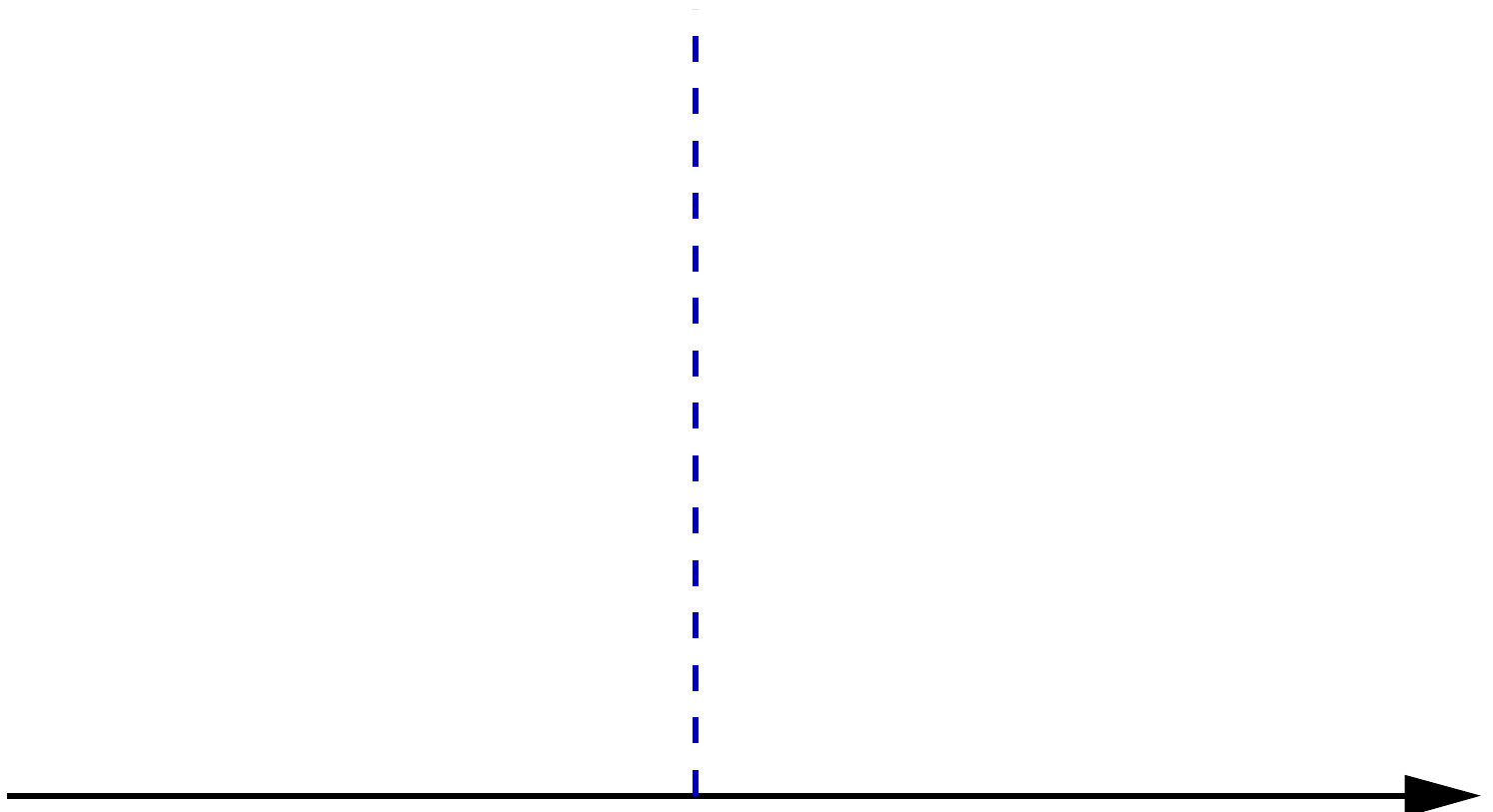}
\put(100,0) {$x$}
\put(0,45) {$\R^{1,3}$}
\put(20,25) {\Huge$\tau_1$}
\put(65,25) {\Huge$\tau_2$}
\put(75,40) {\Large$\tau = \f{\theta}{2\pi} + \f{4\pi i}{g_\text{YM}^2}$}
\end{overpic}
\caption{\label{interface}The complexified coupling constant of ${\cal N}=4$ SYM jumps from one value to another on a 2+1 dimensional hypersurface. As a result a 2+1 dimensional conformal symmetry is retained.}
\end{figure}

Interfaces and defects play a prominent role in many physical system. Phenomena ranging from defects in lattice models to D-branes in string theory are but few examples. Our focus here will be on co-dimension one interfaces where the coupling constants of the theory vary in a direction orthogonal to the interface, but no new dynamical degrees of freedom are introduced \cite{Bak:2003jk}. Such position dependent coupling clearly break the four-dimensional Poincar\'e symmetry to a three-dimensional Poincar\'e invariance. If the original four-dimensional theory is a conformal field theory (CFT), the conformal invariance is similarly also broken. Three-dimensional conformal symmetry can be retained by letting the coupling constants jump from one value to another on an infinitesimally thin surface (see figure \ref{interface}).
% \begin{figure}
% \centering
% \includegraphics[width =0.6\textwidth]{interface.png}
% \caption{\label{interface}The complexified coupling constant of ${\cal N}=4$ SYM jumps from one value to another on a 2+1 dimensional hypersurface. As a result a 2+1 dimensional conformal symmetry is retained.}
% \end{figure} 
These configurations are called Janus interfaces. In this contribution I review the constructions of holographic duals to such Janus interfaces where the parent theory is either four-dimensional ${\cal N}=4$ supersymmetric Yang-Mills (SYM) or a class of ${\cal N}=1$ SFCTs with a marginal coupling. 

The original holographic Janus is an $\SO(6)$ invariant interface in ${\cal N}=4$ SYM \cite{Bak:2003jk}. It breaks all supersymmetries and is realized as a dilatonic deformation of the well-known AdS$_5 \times S^5$ background of type IIB supergravity. It was quickly understood that Janus interfaces in ${\cal N}=4$ could also preserve some ammount of supersymmetry \cite{Clark:2004sb,DHoker:2006qeo}, but then the $\SO(6)$ R-symmetry is also broken. The classification of supersymmetric Janus interfaces in ${\cal N}=4$ was carried out in \cite{DHoker:2006qeo} and is summarized in table \ref{Jclass}.
\begin{table}[h]
\renewcommand{\arraystretch}{1.0}
\centering
\begin{tabular}{@{\extracolsep{15 pt}} cccc}
\hline
\noalign{\smallskip}
$\N$ & Superalgebra & R-symmetry & Commutant \\
\noalign{\smallskip}
\hline
\noalign{\smallskip}
4 & $\OSp(4|4,{\R})$ & $\SU(2)\times\SU(2)$ & \\[4 pt]
2 & $\OSp(2|4,{\R})$ & $\U(1)$ & $\SU(2)$ \\[4 pt]
1 & $\OSp(1|4,{\R})$ & &$\SU(3)$\\ 
%0 &  & &$\SU(4)$
%\noalign{\smallskip}
\hline
\end{tabular}
\caption{\label{Jclass}Possible symmetry superalgebras of the three-dimensional superconformal Janus interfaces inside ${\cal N}=4$ SYM. R-symmetry of the Janus interface must be embedded in the $\SO(6)$ R-symmetry of the parent theory while its commutant when left unbroken, realizes a flavor symmetry of the interface.}
\end{table}
Supersymmetry allows for a more controlled study of the Janus interfaces. In particular $\N=4$ superconformal interfaces were understood to give rise to interesting SCFTs in three dimensions \cite{Gaiotto:2008sa,Gaiotto:2008sd,Gaiotto:2008ak}. Similar study was carried out for $\N=2$ interfaces in \cite{Hashimoto:2014vpa,Hashimoto:2014nwa}. 

In this contribution I focus on the holographic duals of the Janus interfaces and their twisted compactification (which I will come to in a moment). The holographic dual to $\N =1$ interface in $\N=4$ SYM with $\SU(3)$ flavor symmetry was constructed directly in type IIB supergravity in \cite{DHoker:2006vfr}. The solution is a deformation of AdS$_5\times S^5$ just like the original non-supersymmetric Janus \cite{Bak:2003jk}. This solution can also be found as an \emph{uplift} \cite{Suh:2011xc} of a five-dimensional solution \cite{Clark:2005te}. Both procedures lead to the same ten-dimensional background, but the five-dimensional approach has the advantage that a priori the field equations reduce to ordinary differential equations as opposed to partial differential equations one usually encounters in ten dimensions. This is because the five-sphere has been consistently truncated away, leaving a supergravity theory in five dimensions. A type IIB dual of the $\N=4$ interface was constructed directly in ten dimensions in \cite{DHoker:2007zhm}. Here I review the five-dimensional construction of the holographic duals to the $\N=2$ and $\N=4$ interfaces in $\N=4$ SYM which were first constructed in \cite{Bobev:2020fon}. These can also be uplifted to ten dimensions using the formulae in \cite{Baguet:2015sma}. I also review \cite{Bobev:2019jbi} which showed that holographic duals to $\N=1$ interfaces in a large class of $\N=1$ SCFTs can be obtained by uplifting the solution of \cite{Clark:2005te} to deformations of AdS$_5\times \text{SE}_5$ for an arbitrary Sasaki-Einstein manifold SE$_5$.

The study of interfaces in four-dimensional SCFTs has lead to new insights into the world of three-dimensional SCFTs. In particular, the 3D $T[U(N)]$ SCFT discovered on the 1/2-BPS interface of $\N=4$ SYM \cite{Gaiotto:2008sa,Gaiotto:2008sd,Gaiotto:2008ak} possesses a $U(N)\times U(N)$ global symmetry which can be gauged leading to a large class of new SCFTs \cite{Assel:2018vtq,Ganor:2014pha,Terashima:2011qi,Gang:2015wya}. These constructions will be denoted as J-fold theories as their holographic dual are simply twisted compactifications of the holographic Janus interfaces \cite{Inverso:2016eet,Assel:2018vtq}. A direct compactification of the direction orthogonal to the interface is obstructed by the dilaton which has a non-trivial profile in that direction. However for a limiting class of Janus solutions for which the dilaton is linear, a Scherk-Schwarz compactification can be performed leading to AdS$_4$ backgrounds in IIB string theory. Instead of the dilaton being periodic around the compactified circle it has an $\SL(2,\Z)$ monodromy. It is worth noting that these AdS$_4$ backgrounds (but not their Janus parent) can be obtained as solutions of a four-dimensional supergravity \cite{Gallerati:2014xra,Guarino:2019oct,Guarino:2020gfe} which are constructed as a truncation of type IIB on $S^5\times \R$.

This contribution is organized as follows, in section \ref{Sec:Janus} I describe the construction of Janus solutions in five-dimensional supergravity and their uplifts to ten dimensions. In section \ref{Sec:Jfolds} we move on to the compactification of a limiting Janus solution to a AdS$_4$ J-fold solution. In section \ref{Sec:Conclusion} I conclude with some remarks and open questions.

\section{Janus}\label{Sec:Janus}
The holographic Janus solutions are constructed in five-dimensional gauged supergravity. When the dual field theory is $\N=4$ SYM in four dimensions the gravity theory in question is the $\N=8$ supergravity with $\SO(6)$ gauge group \cite{Gunaydin:1984qu,Gunaydin:1985cu,Pernici:1985ju}. The $\SO(6)$ gauge symmetry is the holographic dual to the $\SO(6)$ R-symmetry of $\N=4$ and as mentioned in the introduction, is broken for the Janus solutions we are after. This theory is obtained as a consistent truncation of type IIB supergravity on $S^5$, and the $\SO(6)$ gauge symmetry of the theory is directly related to the isometry group of the five-sphere. The symmetries of the Janus interfaces, listed in table \ref{Jclass}, are an important tool in managing the complicated five-dimensional supergravity theory. Working directly with its 42 scalar fields, 15 vectors and 12 tensor-fields quickly leads to equations which are unmanageable. First we use the fact that all Janus interfaces we study are conformal, preserving $\SO(3,2)$ conformal group in three dimensions. This means that the metric can be written in a domain wall form
\be\label{sugrametric}
\dd s_5^2 = \dd r^2 + \e^{2A}\dd s^2_{\text{AdS}_4}\,,
\ee
where the function $A$ only depends on the radial coordinate $r$. The metric on AdS$_4$ is chosen to have unit radius and its isometries realize the conformal symmetry. Preserving this form of the metric means that all vector and tensor fields must vanish. The Janus interfaces are therefore solution of scalar-metric theory where all scalars as well as the metric only depend on the coordinate $r$. Even with this simplification our remaining task is still seemingly out of reach. This is because the scalar fields are 42, parametrizing the scalar coset $E_{6(6)}/\USp(8)$, and the scalar potential is a complicated function of those 42 fields. In fact, the critical points of the potential have only recently been classified using machine learning techniques \cite{Krishnan:2020sfg,Bobev:2020ttg}. Fortunately, all Janus interfaces can possess continuous global symmetry as a combination of R-symmetry and flavor symmetry. On the supergravity side, this implies that all scalar fields which are charged with respect to this continuous symmetry must vanish. In this way we are able to greatly simplify the scalar sector of the theory. Instead of working with the full coset $E_{6(6)}/\USp(8)$ we are left with the much simpler cosets listed in table \ref{cosets}.
\begin{table}[h]
\renewcommand{\arraystretch}{1.0}
\centering
\begin{tabular}{@{\extracolsep{15 pt}} ccc}
\hline
\noalign{\smallskip}
$\N$ & Scalar coset & dimension  \\
\noalign{\smallskip}
\hline
\noalign{\smallskip}
4 & $\frac{\SL(3,\R)}{\SO(3)}$ & 5 \\[4 pt]
2 & $\R_+\times\frac{\SO(3,2)}{\SO(3)\times \SO(2)}$ & 7 \\[4 pt]
1 & $\frac{\SU(2,1)}{\U(2)}$ & 4\\ 
\hline
\end{tabular}
\caption{\label{cosets}Scalar cosets and their dimension for the three Janus interfaces considered.}
\end{table}
It should be noted that for $\N=2$ and $\N=1$ supersymmetric Janus interfaces, we assume the maximum amount of flavor symmetry. It would be interesting to search for Janus solutions where this restriction is relaxed, but then the scalar sector of the supergravity theory is bigger than considered here and solutions are harder to obtain.

The five-dimensional $\SO(6)$ gauged supergravity exhibits a \emph{global} $\SL(2,\R)_S$ symmetry which corresponds directly to the $\SL(2,\R)$ symmetry of type IIB supergravity. This implies in particular that the full five-dimensional potential does not depend on two scalar fields $(\varphi,c)$. These scalar fields are usually called the axion and dilaton as they are closely related to the type IIB axion and dilaton. It is important to note however that they are not just a direct dimensional reduction of the type IIB fields. I will make this distinction clear by  using capital letters $(\Phi,C)$ to denote the ten-dimensional dilaton and axion. The five-dimensional axion and dilaton parametrize the Euclidean disc $\SL(2,\R)/\SO(2)$ and are uncharged with respect to $\SO(6)$. They are therefore a part of all three scalar truncations mentioned above,
\be
\frac{\SL(2,\R)}{\SO(2)}\subset {\cal M} \subset \frac{E_{6(6)}}{\USp(8)}\,,
\ee
where ${\cal M}$ is one of the coset manifolds in table \ref{cosets}. It turns out that by parametrizing the scalar manifold in a particular way, one can choose a gauge for the $\SL(2,\R)$ symmetry such that two scalars are eliminated completely.\footnote{This was shown in \cite{Bobev:2020fon} using the BPS equations.} Schematically the coset element $U\in {\cal M}$ is written as a product
\be
U = V \cdot U_{\SL(2)}\,,
\ee
where $U_{\SL(2)}$ parametrizes the full $\SL(2,\R)$ and not only its non-compact part. The matrix $V$ parametrizes the rest of the manifold ${\cal M}$. The $\SL(2,\R)_S$ symmetry acts on $U$ by a left multiplication and so essentially acts on $U_{\SL(2)}$ in this parametrization. As mentioned, the $\SL(2,\R)_S$ symmetry can be utilized to simplify $U_{\SL(2)}$ such that it only involves a single scalar -- the dilaton. A general background, in an arbitrary $\SL(2,\R)_S$ gauge can be obtained by rotating the solution we find using $\SL(2,\R)_S$ transformation.

The final step in obtaining the Janus backgrounds is to impose supersymmetry in line with the supersymmetry preserved by the Janus interfaces in table \ref{Jclass}. This is done by making sure that fermion variations of the $\N=8$ supergravity theory vanish. For flat supersymmetric domain walls we often encounter BPS equations of the form 
\be\label{flatBPS}
A'\sim W\,,\quad \phi' \sim \partial_\phi W\,,
\ee
where $W$ is the superpotential of the truncation, $\phi$ denotes a collection of scalar fields in the truncation and prime is the derivative with respect to $r$. For the Janus interfaces we are after the domain walls are curved and therefore the BPS equations are more complicated. For example the metric function $A$ now satisfies
\be\label{Aprimeeq}
(A')^2 = \f19 |W|^2 - \e^{-2A}\,,
\ee
where the absolute value appears because the superpotential of our truncations are usually complex. In order to go further we must analyze each Janus background in turn. Due to space constraints I will be brief but further details can be found in \cite{Bobev:2020fon}.

\subsection{$\N=4$ Janus}
\begin{figure}[h]\centering
\includegraphics[width=0.6\textwidth]{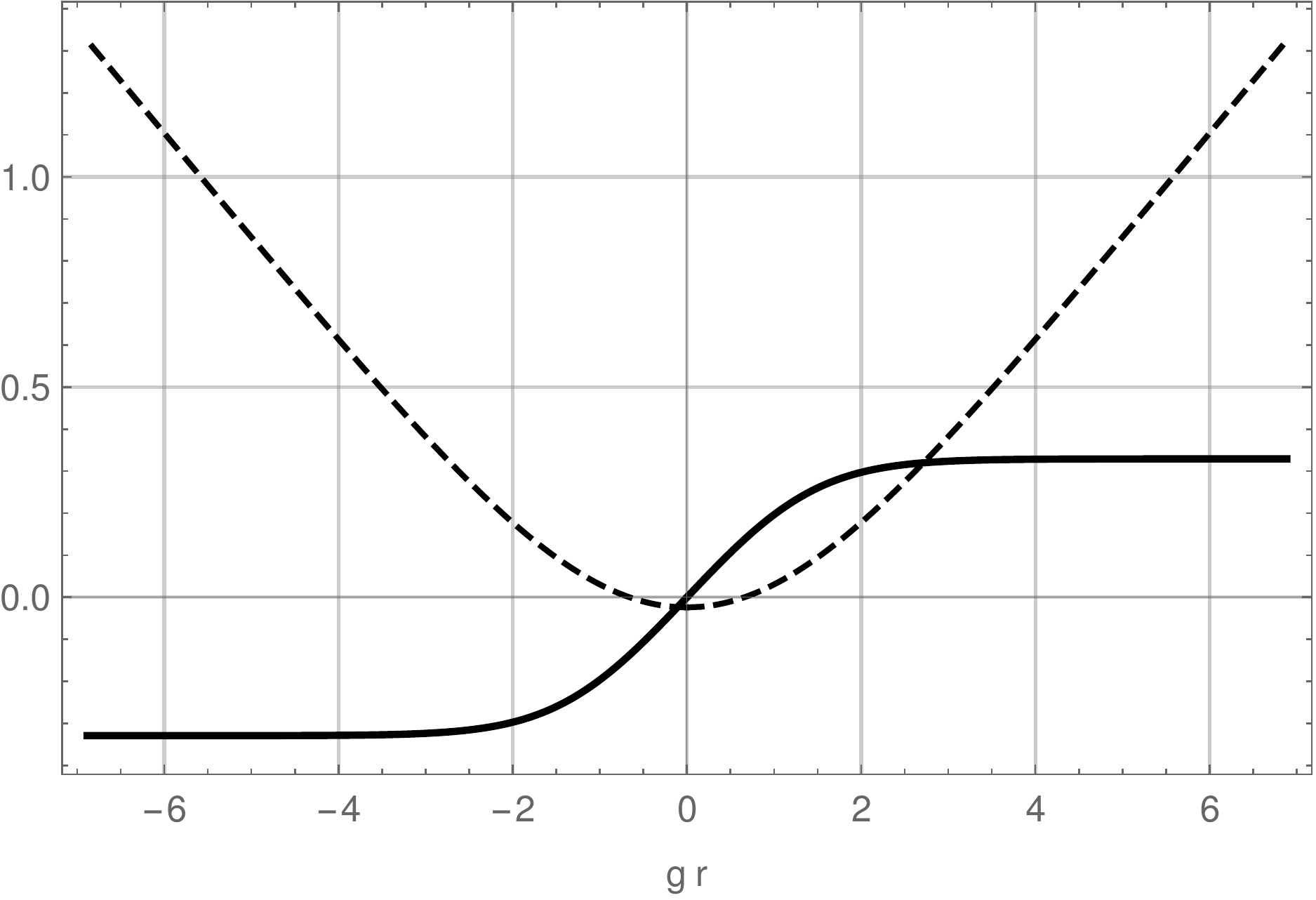}
\caption{\label{N4plot}A plot of the $\N=4$ Janus solution with ${\cal I}=1/3$. Solid line shows the dilaton $\varphi-\varphi_0$ and the dashed line shows the metric function $(1/4)\log X$.}
\end{figure}

The scalar truncation for the $\N=4$ Janus features the superpotential
\be
W = \f{-3g}{2}\Big(\cosh2\alpha\,\cosh2\chi - i\sinh2\alpha\,\sinh2\chi\Big)\,,
\ee
where $\alpha$ and $\chi$ are the two scalars in the truncation in addition to the dilaton $\varphi$ (after we have fixed a gauge to eliminate two scalars). Here $g$ is the gauge coupling constant in the $\N=8$ theory and determines the length scale of the AdS$_5$ vacuum $L=2/g$. The BPS equations, in addition to \eqref{Aprimeeq} are
\be\label{N4BPS}
\begin{split}
35(\alpha'-\varphi')^2 &= |\partial_\alpha W|^2\,,\\
24(\alpha'-\varphi')(\chi') &= \sinh 4\chi\, \Re(W\partial_\alpha W)\,,\\
24(\alpha'-\varphi')(\varphi') &= \tanh 4\chi\, \Im(W\partial_\alpha W)\,,\\
18(\alpha'-\varphi')\e^{-A} &=\Im(W\partial_\alpha \overline{W})\,.
\end{split}
\ee
And we notice the clear difference with the flat domain wall BPS equations \eqref{flatBPS}. Note that due to the last equation in \eqref{N4BPS} we can solve for one of the functions $(A,\chi,\alpha)$ algebraically. In fact the system of BPS equations exhibits a further integral of motion ${\cal I}>0$ which means that we can write
\be\label{N4Xdef}
\cosh 4\alpha = \f{2X^2 +{\cal I}}{2\sqrt{X^4+{\cal I}X}}\,,\quad \sinh4\chi = \sqrt{{\cal I}X^{-3}}\,,\quad \e^{2A} = \f{4}{g^2} X\,.
\ee
I introduced a new function $X$, which determines the full background. A differential equation for $X$ can be obtained by inserting these three relations into \eqref{Aprimeeq} which yields
\be\label{N4classicalmech}
\f{4}{g^2}(X')^2 + V_\text{eff} = 0\,,\quad V_\text{eff}=4X(1-X)-{\cal I}\,.
\ee
The BPS equations are therefore reduced to the classical mechanics problem of a one-dimensional particle with zero energy in the potential specified by $V_\text{eff}$. The holographic Janus solutions should be asymptotically AdS$_5$ reflecting the fact that far away from the Janus interface all experiments should yield the same result as if we were in the conformal vacuum of $\N=4$ SYM. The AdS asymptotics is reached as $X\to \infty $ in this language. We should reach these asymptotics on both sides of the interface and so we want our classical particle described by $X$ to come in from infinity and bounce of the potential where it turns around and heads back to $+\infty$. This implies that the constant ${\cal I}$ is bounded from above by ${\cal I}\le 1$. The analytic solution to \eqref{N4classicalmech} is given by
\be
X = \f12 \Big(1+\sqrt{1-{\cal I}}\cosh(gr -g r_\text{tp})\Big)\,,
\ee
which explicitly shows the bouncing behavior I just described. At $r=r_\text{tp}$ the velocity $X'$ changes from being negative to positive. I will choose coordinates such that $r_\text{tp}=0$. Finally, in order to fully specify the background we must solve the BPS equation for the dilaton. A straight-forward manipulation of the BPS equations together with \eqref{N4Xdef} results in the solution for $\varphi$ as a function of $X$
\be\label{N4dilaton}
\varphi(X) = \varphi_0 \pm \int_{X_\text{tp}}^X \f{3\sqrt{{\cal I}}({\cal I} + 2 x^2)}{8x({\cal I} + x^3)} \f{\dd x}{\sqrt{-V_\text{eff}}}\,,
\ee
where $X_\text{tp} = X(r_\text{tp})$. The sign choice in \eqref{N4dilaton} is directly correlated with the branch of the square root in \eqref{N4classicalmech} and we must switch a branch when passing through the turning point to get a regular solution. A sample solution is plotted in figure \ref{N4plot}.

This solution can be uplifted to a ten-dimensional solution of type IIB supergravity as explicitly demonstrated in \cite{Bobev:2020fon}. There it was also shown that the uplifted solution reproduces the ten-dimensional Janus backgrounds found previously in \cite{DHoker:2007zhm}.

\subsection{$\N=2$ Janus}
The $\N=2$ Janus is a priori significantly more complicated than the $\N=4$ one. This is because the scalar truncation is bigger. Surprisingly, however, we manage to bring the BPS equations essentially to the same form as for the $\N=4$ Janus. The only modification is the effective potential which now is more complicated. The superpotential is
\be
W = -\f{g}{2}\e^{-4\alpha}\Big(2\e^{6\alpha}\cosh2\chi + \cosh2\lambda - i \sinh 2\lambda \sinh2\chi\Big)\,,
\ee
where $(\alpha,\chi,\lambda)$ are the three scalars in the truncation in addition to the dilaton. As mentioned above, the scalar coset in this case is seven-dimensional. Two scalars are eliminated by choosing a gauge for the $\SL(2,\R)_S$ symmetry and one more is eliminated by choosing a gauge for a broken $\U(1)\subset\SO(6)$ symmetry \cite{Bobev:2020fon}. It should be noted that even though I reuse the name for some of the scalars, their origin within the $\N=8$ supergravity theory is different and should not be confused. Just as in the $\N=4$ case, the BPS equations can be expressed neatly in terms of the superpotential. Due to space constraints I will omit them here. Remarkably the BPS equations again exhibit a number of integrals of motion which greatly simplifies the analysis. In fact, both the metric as well as the scalars $\chi$ and $\lambda$ can be expressed in terms of $\alpha$:
\be
\e^{6\alpha}\cosh2\lambda\,\cosh2\chi = 1\,,\quad {\cal I} = -\f{32\e^{6\alpha}\sinh^36\alpha}{\sinh^42\chi}\,,\quad \e^{-2A} = \f{g^2}{2\sqrt{{\cal I}}}\sqrt{-2\e^{2\alpha}\sinh6\alpha}\,,
\ee
where ${\cal I}>0$ is an integration constant which plays exactly the same role as the integration constant for $\N=4$ Janus. The classical mechanics problem is now stated in terms of 
\be
X = -2\e^{6\alpha}\sinh6\alpha\,,
\ee
for which the effective potential is
\be
V_\text{eff} = - \f{16(1-X)^{1/3} X^2}{\sqrt{{\cal I}}}\Big(\sqrt{I}-2\sqrt{X(1-X)}\Big)\,.
\ee
The remaining analysis closely follows the $\N=4$ case with two exceptions. First the AdS asymptotic region is now located at $X\to0$, and second the equation for $X$ can not be analytically solved. Numerically, however, the solution is straightforward to obtain when we express the radial coordinate as a function of $X$ 
\be
r(X) = r_\text{tp} \pm \int_{X_\text{tp}}^X \f{2\dd x}{g\sqrt{-V_\text{eff}(x)}}\,,
\ee
where $X_\text{tp}$ is the turning point and is obtained as the first zero of the effective potential. Regular Janus solutions (for which there is a turning point) restrict the range ${\cal I}\le 1$. The radial location of the turning point, $r_\text{tp}$, can be chosen to vanish for simplicity. Next the dilaton is obtained in much the same way as before via the equation
\be
\varphi(X) = \varphi_0 \pm \int_{X_\text{tp}}^X \f{(3-2x)x}{\sqrt{{\cal I}}(1-x)^{4/3} + 2(1-x)^{5/6}x^{3/2}} \f{\dd x}{\sqrt{-V_\text{eff}}}\,.
\ee

Once again I refer to \cite{Bobev:2020fon} for much more detailed analysis as well as the uplift to ten dimensions of these solutions.

\subsection{$\N=1$ Janus}
Finally we review the construction of $\N=1$ Janus solutions in five-dimensional supergravity which was first carried out in \cite{Clark:2005te}. The solutions there can be uplifted to ten dimensions \cite{Suh:2011xc} resulting in Janus solutions which were independently constructed directly in ten dimensions \cite{DHoker:2006vfr}. 

It was realized in \cite{Bobev:2019jbi} that the scalar truncation relevant for holographic duals to ${\cal N}=1$ interfaces in $\N=4$ SYM is a part of minimal $\N=2$ gauged supergravity coupled to a single hypermultiplet. From this perspective the $\N=1$ Janus we now review is a universal solution dual to an interface of any holographic $\N=1$ SCFT in four dimensions with a marginal coupling (the latter being dual to the hypermultiplet). We will return to this interpretation at the end of the section.

My treatment follows closely \cite{Bobev:2020fon} and \cite{Bobev:2019jbi} which uses a different parametrization of the scalar manifold than in  \cite{Clark:2005te}. This means that a gauge can be chosen such that only two scalars $(\chi,\varphi)$ out of four scalars have a non-trivial profile as explained above. The superpotential is
\be
W = -\f{3g}{2}\cosh^2\chi\,.
\ee
As before, the BPS equations exhibit constants of motion which essentially reduce the problem to a classical mechanics problem of a single particle. Here
\be
{\cal I} = \f{9g^2}{5^{5/3}}\e^{2A}\sinh^{2/3}\chi\,,\quad X = -\f13 \log\sinh \chi
\ee
and the effective potential experienced by $X$ is
\be
V_\text{eff} = 4\e^{-2X}\Big(\f{9}{5^{5/3}{\cal I}}-\e^{-4X}\cosh^23X\Big)\,.
\ee
The asymptotic AdS region is located at $X\to\infty$, and the requirement to have a turning point forces $0<{\cal I}\le 1$.
The classical mechanics problem is not analytically solvable but $r(X)$ can be written as a simple integral as for the $\N=2$ Janus. The dilaton is then also found via a similar integral
\be
\varphi = \varphi_0 \pm \int_{X_\text{tp}}^X \f{9\e^{-x}}{5^{5/6}\sqrt{{\cal I}}\cosh 3x}\f{\dd x}{\sqrt{-V_\text{eff}(x)}}\,.
\ee

As before, these Janus configurations can be explicitly embedded as solution to type IIB supergravity on $S^5$\cite{Suh:2011xc,Bobev:2020fon}. The ten dimensional metric takes the simple form 
\be\label{N110Dmetric}
\dd s_{10}^2 = \cosh\chi \,\,\dd s_5^2 + \f{4}{g^2} \Big( \frac{\dd s_{\mathbf{CP}^2}^2}{\cosh\chi} + \cosh\chi\,\, \zeta^2 \Big)\,,
\ee
where $\dd s_5^2$ is the five-dimensional supergravity metric \eqref{sugrametric} and $\zeta=\dd \phi +\sigma$ is the $\U(1)$ bundle over the $\mathbf{CP}^2$ base. The K\"ahler form $J$ is the exterior derivative of $\zeta$. The $\SU(3)$ symmetry of the Janus is realized as the isometry group of $\mathbf{CP}^2$. For vanishing $\chi$ (as $X\to\infty$) the terms in the bracket in \eqref{N110Dmetric} reduce to the round metric on $S^5$
\be
\dd s_{S^5}^2 = \dd s_{\mathbf{CP}^2}^2 + \zeta^2\,,
\ee
with unit radius. The remaining ten-dimensional fields are all determined either by five-dimensional supergravity data or in terms of geometric forms on $\mathbf{CP}^2$. 

Arbitrary five-dimensional Sasaki-Einstein manifolds can be used to construct AdS$_5$ solution of type IIB supergravity in complete analogy with AdS$_5\times S^5$.  Here the metric on the five-sphere is directly replaced by the Einstein metric on the Sasaki-Einstein manifold. The dual field theory is in this case the theory on D3 branes probing the singularity at the apex of the cone over that Sasaki-Einstein manifold. These are strongly coupled, often non-Lagrangian, SCFTs that all share the feature of having at least one marginal deformation (dual to the ten-dimensional axion-dilaton). As explained in \cite{Bobev:2019jbi}, the five-dimensional Janus solution with $\N=1$ supersymmetry just reviewed can be uplifted to type IIB supergravity not only on $S^5$, but on any of those Sasaki-Einstein manifolds. Explicitly we use the fact that metric on Sasaki-Einstein manifolds can be written as $\U(1)$ fibration over a K\"ahler-Einstein base. The metric \eqref{N110Dmetric} is then only modified in that $\dd s_{\mathbf{CP}^2}^2$ is replaced by the metric on the K\"ahler-Einstein base in question and $\zeta$ is replaced by the corresponding fiber. The remaining supergravity fields receive similar modest adjustments. This shows that any holographic $\N=1$ SCFT with a marginal coupling can be deformed to include a Janus interface just like $\N=4$ SYM.

\section{J-folds}\label{Sec:Jfolds}
For the three classes of Janus solutions we were able to reduce the BPS equations to that of a classical mechanics system of a particle in a potential with zero energy. The Janus solution correspond to a trajectory for the particle where it bounces of the potential much like an unbounded null geodesic in the Schwarzschild geometry. In all cases, however, the effective potential $V_\text{eff}$ has a critical point with zero potential energy for ${\cal I}=1$. There is therefore an exact solution to the BPS equations where instead of the bouncing dynamics, the particle simply sits stationary on the critical point. This is analogous to the unstable photon ring in the Schwarzschild problem. In the current case, these stationary solutions are not unstable. One way of understaning this is the unstable direction appears as a tachyon in AdS$_4$. This tachyon has a mass that is above the four-dimensional BF bound. This class of solutions is a particularly simple limiting case of the Janus solutions I have reviewed. Since $X$ is constant, the metric is in a product form $\R\times \text{AdS}_4$ and all scalars except the dilaton are constant. The dilaton is linear. It is important to note that these solutions can not be interpreted as  Janus solutions unless we glue to this linear region a dynamical part where the classical mechanics particle arrives from the asymptotic region and leaves to it again. Without these asymptotic pieces, the solutions are a priori singular (as the dilaton diverges as $r\to\pm\infty$) and should be dismissed. If it were not for the dilaton, we could compactify this solution and view it as a new AdS$_4\times S^1$ background in type IIB supergravity. 

Even if the dilaton is not constant or periodic, there is a way to compactify the radial direction using the $\SL(2,\R)_S$ symmetry of the five-dimensional supergravity \cite{Inverso:2016eet,Assel:2018vtq}. In short, we follow an old procedure of compactifying supergravity solutions using global symmetries as introduced by Scherk and Schwarz \cite{Scherk:1979zr}. This introduces a monodromy for the dilaton as we go around the compactified direction. These compactified solutions are called J-fold backgrounds as they are essentially S-folds originating from a Janus construction. Since the symmetry group of type IIB \emph{string theory} is $\SL(2,\Z)$ and not $\SL(2,\R)$ the radius of compactification is quantized which is an important fact when making holographic predictions. Now that we have managed to compactify the $\R$ direction we have obtained a bone-fide AdS$_4$ background of type IIB string theory. These should be dual to three-dimensional $\N=1,2,4$ SCFTs. In order to make connection to those SCFTs, we compute the $S^3$ free energy using holography, the results are listed in table \ref{freee}.
\begin{table}[h]
\renewcommand{\arraystretch}{1.0}
\centering
\begin{tabular}{@{\extracolsep{15 pt}} cc}
\hline
\noalign{\smallskip}
$\N$ & Free energy (${\cal F}_{S^3}$)  \\
\noalign{\smallskip}
\hline
\noalign{\smallskip}
4 & $\f{N^2}{2}\text{arccosh}(n/2)$ \\[4 pt]
2 & $\f{N^2}{2}\text{arccosh}(n/2)$ \\[4 pt]
1 & $\f{5^{5/2}N^2}{4 \times 3^3}\text{arccosh}(n/2)$\\ 
\hline
\end{tabular}
\caption{\label{freee}The three-sphere free energy of the J-fold SCFTs associated with $\N=4$ SYM, computed using holography. The integer $n=2,3,4,\cdots$ is related to the radius of compactification of the $S^1$.}
\end{table}
For the $\N=4$ J-fold, the dual field theory was identified and an independent computation of the free energy using localization revealed a perfect match with the supergravity prediction \cite{Assel:2018vtq}. Similar success has not been acheived for less supersymmetric J-folds but they should explain the surprising fact that $\N=4$ and $\N=2$ J-folds have the same free energy. Possible explanations for this fact are offered in \cite{Bobev:2020fon}. As expected, the $\N=1$ J-fold can be embedded as a solution of type IIB on any five-dimensional Sasaki-Einstein manifold and the free energy can be computed to be \cite{Bobev:2019jbi}
\be
{\cal F}_{S^3} =\f{5^{5/2}}{3^3} a_{4d} ~\text{arccosh}(n/2)\,,
\ee
where $a_{4d}$ it he central charge of the parent four-dimensional $\N=1$ SCFT.

Finally, it should be noted that the J-fold backgrounds can be constructed directly without first finding the full Janus solution as we have done here. In this case a natural starting point is a four-dimensional supergravity obtained as a consistent truncation of type IIB on $S^5\times\R$. This approach was taken in \cite{Gallerati:2014xra,Guarino:2019oct,Guarino:2020gfe} leading to the J-fold solutions discussed here as well as some novel ones I have not mentioned.

\section{Conclusion}\label{Sec:Conclusion}
In this contribution I have sketched the construction of holographic Janus backgrounds in type IIB supergravity using a consistent truncation to five-dimensional supergravity. Remarkably the BPS equations exhibit the same structure regardless the amount of supersymmetry and a full classification of solutions is possible. An obvious future direction is to relax the global symmetry imposed for the less supersymmetric Januses. The scalar truncation is then expected to be considerably more complicated. It is still possible that the structure we encountered for the Janus reviewed here could show itself again leading to solvable equations. Another ambitious, but worthwhile direction is to attempt to classify the J-fold backgrounds in type IIB on $S^5\times S^1$. This is essentially equivalent to classifying the vacua of the four-dimensional supergravity  used in \cite{Gallerati:2014xra,Guarino:2019oct,Guarino:2020gfe}. As we have seen in recent months, the techniques of machine learning could be invaluable in this direction \cite{Comsa:2019rcz,Bobev:2019dik,Krishnan:2020sfg,Bobev:2020ttg}. Another direction is to analyze non-conformal Janus solutions in supergravity. As far as I am aware this is a largely unexplored territory perhaps not unexpectedly since it requires dealing with PDEs rather than ODEs even in five dimensions.

%%%%%%%%%%%%%%%%%%%%%%%%%%%%%%%%%%
\section*{Acknowledgements}
It is a pleasure thank the organisers of ``School and Workshops on Elementary Particle Physics and Gravity'' (CORFU2019). I am also deeply grateful to my collaborators Nikolay Bobev, Krzysztof Pilch, Minwoo Suh, and Jesse van Muiden.  I am a Postdoctoral Fellow of the Research Foundation - Flanders (FWO). I am also supported by the KU Leuven C1 grant ZKD1118 C16/16/005.
%%%%%%%%%%%%%%%%%%%%%%%%%%%%%%%%%%

\bibliographystyle{JHEP}
\bibliography{references}

%%% --- DOCUMENT END --- %%%
\end{document}